\documentclass[journal]{IEEEtran}
\usepackage{xcolor}
\usepackage{amsmath}
\usepackage{cite}
\usepackage{url}
\usepackage{algorithm,algorithmic}
\usepackage[keeplastbox]{flushend}
\ifCLASSINFOpdf
   \usepackage[pdftex]{graphicx}
\else

\fi

\begin{document}
\title{A Blind Signal Separation Algorithm \\for Energy Detection of Dynamic PU Signals}
\author{Jakub~Nikonowicz, Aamir~Mahmood,~\IEEEmembership{Senior~Member,~IEEE} and~Mikael~Gidlund,~\IEEEmembership{Senior~Member,~IEEE}
\thanks{J. Nikonowicz is with the Faculty of Computing and Telecommunications, Pozna{\'n} University of Technology, 61-131 Pozna{\'n}, Poland, e-mail: jakub.nikonowicz@put.poznan.pl.}
\thanks{A. Mahmood and M. Gidlund are with the Department of Information Systems and Technology, Mid Sweden University, 851 70 Sundsvall, Sweden.}
\vspace{-12pt}
}


\maketitle

\begin{abstract}

Energy detection process for enabling opportunistic spectrum access in dynamic primary user (PU) scenarios, where PU changes state from active to inactive at random time instances, requires estimation of several parameters ranging from noise variance and signal-to-noise ratio (SNR) to instantaneous and average PU activity. A prerequisite to parameter estimation is an accurate extraction of the signal and noise samples in a received signal time frame. 
In this letter, we propose a low-complexity and accurate signal separation algorithm as compared to well-known methods, which is also blind to the PU activity distribution. 
The proposed algorithm is analyzed in a semi-experimental simulation setup for its accuracy and time complexity in recognizing signal and noise samples, and its use in channel occupancy estimation, under varying occupancy and SNR of the PU signal. The results confirm its suitability for acquiring the necessary information on the dynamic behavior of PU, which is otherwise assumed to be known in the literature.
\end{abstract}


%
\IEEEpeerreviewmaketitle

\vspace{-10pt}
\section{Introduction}
\label{sec:intro}

\IEEEPARstart{I}{n} cognitive radios, detecting white spaces, and determining channel occupancy in a dynamic radio environment is essential for opportunistic access to radio resources. The simplest and widely used method of assessing the availability of the radio resources is energy detection, for which efficiency in terms of probability of detection and probability of false alarm is analyzed extensively in the literature (see \cite{Urkowitz1967, Sharma2015, Ali2017}). However, these studies consider the context of a static primary user (PU) signal, which is either active or inactive within the entire detection period. In realistic scenarios, however, the PU signal can dynamically switch between active and inactive states, while detection is in progress. The detection of discontinuous PU is considered in \cite{Penna2011, Saad2016, MacDonald2017, Duzenli2019} by redesigning the detection algorithms, in which, however, signal, noise, and PU traffic parameters are assumed to be known \textit{a priori} while the methods to obtain/estimate these necessary parameters are not described. 

Many studies show that any uncertainty in estimating parameters of the received signal seriously limits the ability of the detector to assign energy to a particular activity state correctly \cite{Mariani2011}. Consequently, the correct operation of the energy detector in a dynamic radio channel requires accurately estimating, a) PU traffic parameters such as the average and current duration of the PU, and its channel occupancy ratio and, b) signal and noise parameters as noise variance and signal-to-noise ratio (SNR). Importantly, the estimation of all these parameters and the ensuing detection performance starts with the accurate splitting of the signal and noise samples in received energy samples.

In this letter, we present a practical algorithm for energy samples separation---marking of signal and noise samples in a received time frame---of a dynamic PU signal. The algorithm uses rank order filtering, earlier studied for signal spectrum analysis only \cite{Taher2014, Agarwal2017, Nikonowicz2019}, for temporal signal analysis by redesigning the signal processing and samples marking processes. We evaluate the algorithm in terms of marking signal samples and complete samples separation with respect to SNR and different PU activity factors, and also examine the execution time of the separation process. 
Besides, its performance is compared with the well-known reference methods in the literature \cite{Toma2020, Iwata2019, Chien2019}, which is then followed by its utility appraisal for channel occupancy estimation.
To assess the accuracy of these operations, a semi-experimental simulation setup of packet-based PU transmission is designed, where the background distortion comes from the radio frequency (RF) noise traces captured with National Instrument software defined radio (SDR), USRP-2900. The proposed solution, with its appealing performance, provides a convenient basis, although not the subject of this article, for parameter estimation in the subsequent detection of intermittent PU signals.

The rest of the article is organized as follows. Section~\ref{sec:system} gives the motivation for samples separation, and Section~\ref{sec:algorithm} describes the proposed separation algorithm. Section~\ref{sec:simulation} explains the simulation methodology and shows numerical results. Finally, Section~\ref{sec:conclusion} gives the concluding remarks.

\section{Motivation for Samples Separation}
\label{sec:system}

\begin{figure}[!t]
	\centering
		\includegraphics[trim=0 0 0 15,clip,width=0.90\linewidth]{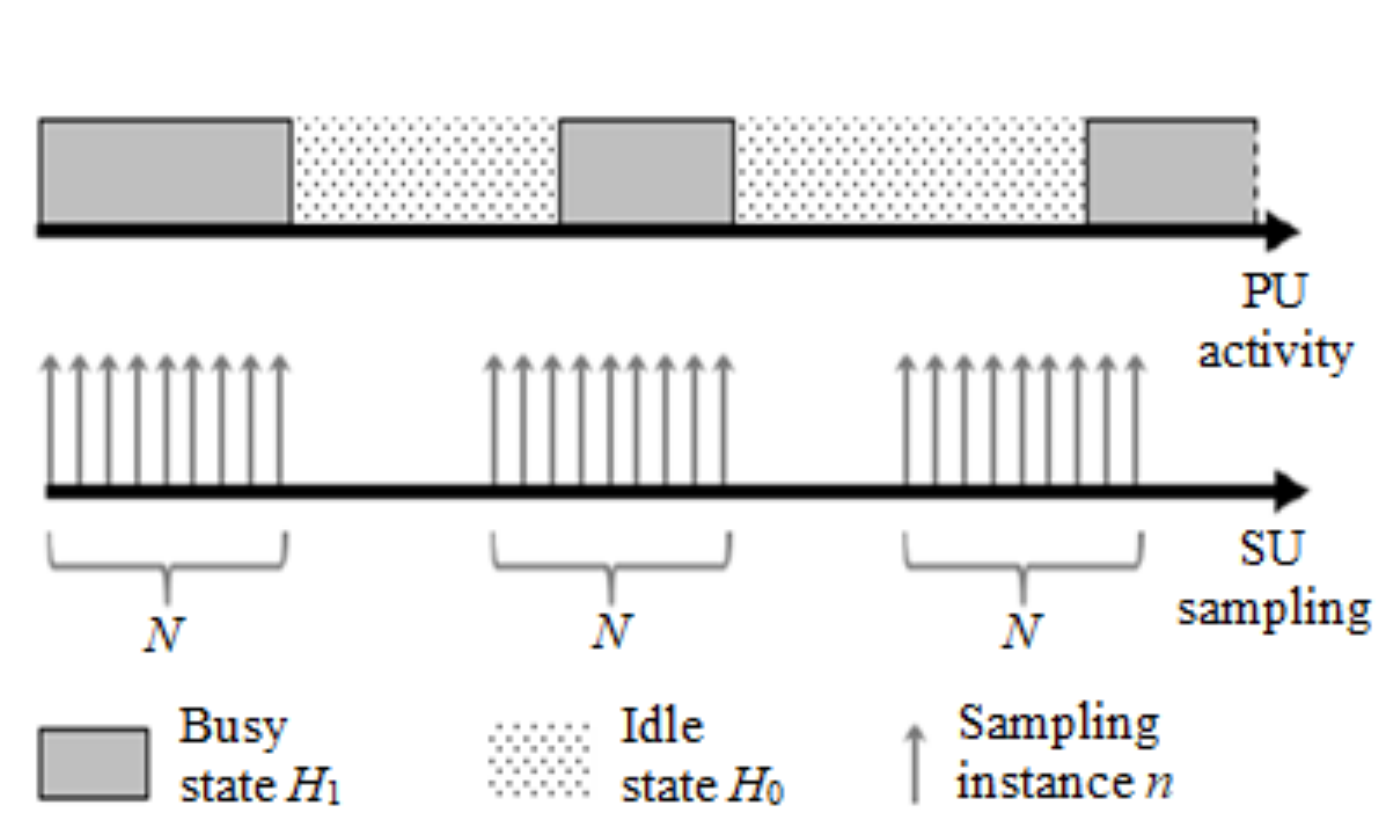}
	\caption{Illustration of a dynamic PU signal activity along with the energy sampling process.}
	\label{fig:transmission}
\end{figure}

To reason the need for samples separation, we restate the energy detection (ED) process for dynamic PU signals. Consider the ED-based sampling of dynamic PU modeled as an alternating renewal process, similar to \cite{Penna2011, Saad2016, MacDonald2017}, and shown in Fig.~\ref{fig:transmission}. At any time instant, PU is either in ON (active) or OFF (idle) state, while the state transition occurs at random time instances, and the state holding times are exponentially distributed with mean $\tau$ and $\mu$, respectively. The energy detector collects signal samples 
$x_{n}, n=1\dots N$ in a detection interval of duration ($T$), which is independent of the PU ON/OFF process. As the signal is sampled at a specific frequency $f_s$, total  number of collected samples are $N = f_{s}{T}$, and $N_{0}$ and $N_{1}$ represent the number of samples corresponding to hypothesis (subject to detection) $H_{0}$ and $H_{1}$. As $N\rightarrow \infty$, normalized occupancy/absence rate ${N_{i}}/{N}$ approaches it average value $p_{i}, i\in \{0,1\}$.

A predominant model to characterize energy detection performance for such dynamic PU scenario, in terms of test statistic ($\beta$), probability of detection ($P_{d}$), and probability of false alarm ($P_{f}$), is as follows \cite{Penna2011}
\begin{equation}
    \beta{}=p_0\frac{1}{N_0}\sum_{n=1}^{N_0}{\left\vert{}w_{n}\right\vert{}}^2+p_1\frac{1}{N_1}\sum_{n=1}^{N_1}{\left\vert{}x_{n}+w_{n}\right\vert{}}^2 ,
\label{eq:beta}
\end{equation}
\begin{equation}
    P_d=Q\left[\sqrt{N}\frac{\frac{\gamma{}}{{\sigma{}}_w^2}-\left(1+p_1\rho{}\right)}{\sqrt{1+p_1\left({\rho{}}^2+2\rho{}\right)}}\right] ,
\label{eq:detprob}
\end{equation}
\begin{equation}
    P_f=Q\left[\sqrt{N}\left(\frac{\gamma{}}{{\sigma{}}_w^2}-1\right)\right] ,
\label{eq:falseprob}
\end{equation}
where $\gamma$ is the decision threshold, $\sigma_{w}^2$ is noise variance, $\rho$ is the SNR. 

\begin{figure}[!t]
	\centering
		\includegraphics[width=1\linewidth]{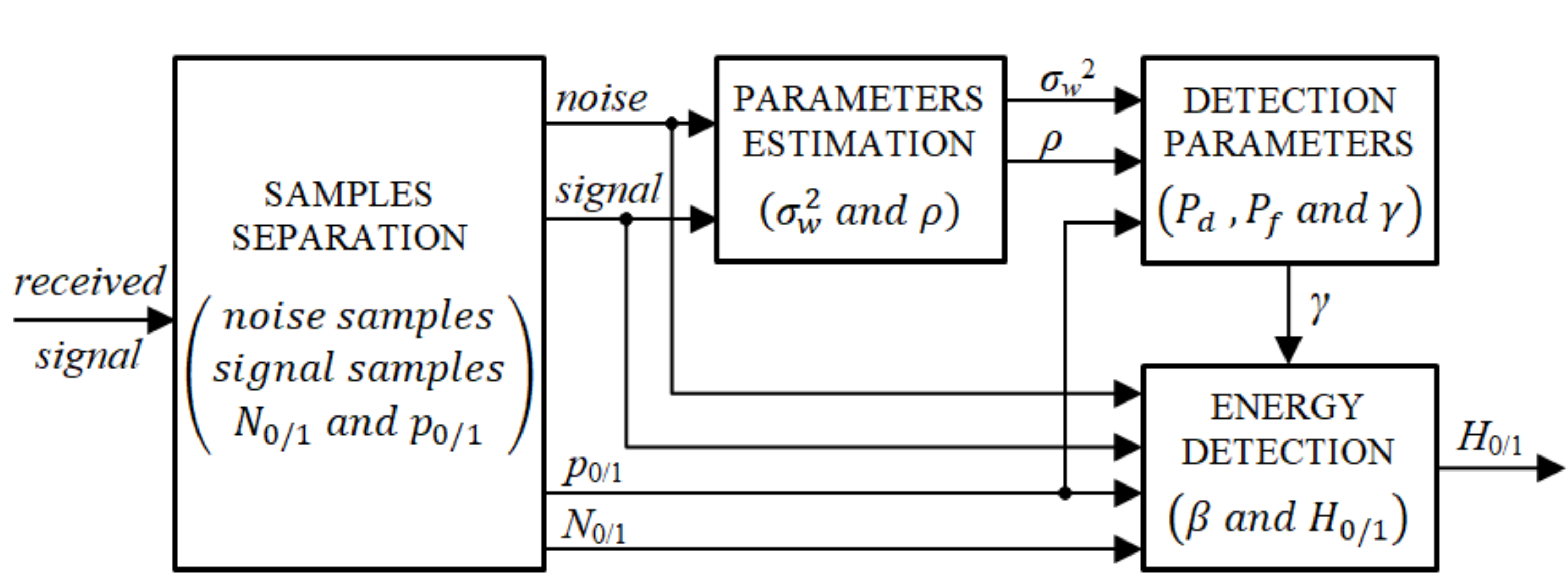}
	\caption{Block diagram of an energy detection process for discontinuous PU signals: source and demand of necessary parameters.}
	\label{fig:system}
\end{figure}

To implement this detection model or any other involving PU transition probabilities (e.g., \cite{MacDonald2017}), several necessary parameters, as noise variance, SNR, instantaneous and mean occupancy/absence rates, are assumed to be known. In practice, the very first step to extract these parameters is samples separation. Fig.~\ref{fig:system} summarizes the different stages of the energy detection process while featuring the source and demand of the necessary parameters at each stage. 
   
In this context, our objective is to develop a samples separation algorithm that is effective in extracting the necessary (detection-related) parameters of bursty PU signal. We assume a specific (exponential) distribution of PU idle/active state; however, the design of the separation algorithm is generic and blind to the PU activity pattern.

\section{Algorithm Design and Description}
\label{sec:algorithm}

In this section, we describe the design of the proposed algorithm, which finds its motivation from rank order filtering (ROF). ROF, a commonly used image processing technique, sorts the input values in ascending order, and selects for the output value encountered at a certain rank order number. The selected input value becomes the output, without any calculation performed on the input values. The two special operations of ROF are \textit{erosion}---equivalent to lowest rank as it returns the minimum of the input set, and \textit{dilation}---equivalent to highest rank as it returns the maximum.
Erosion and dilation, besides being useful in image processing, can also be effectively used in impulse noise reduction and noise power estimation, as demonstrated in \cite{Taher2014, Agarwal2017, Nikonowicz2019}. These studies iteratively increase the size of the filters used on the power spectrum samples. By filtration, the peak values of the spectrum are reduced until the difference in the noise floor achieved in $i$--th and $(i+1)$--th iteration falls below a predetermined threshold value. Although effective, the algorithms in \cite{Taher2014, Agarwal2017, Nikonowicz2019} are only dedicated to estimating spectrum parameters and are burdened with the following disadvantages that limit their usefulness for time-domain analysis of dynamic PU behavior: 
\begin{itemize}
    \item The selection of an appropriate threshold for the noise floor difference is problematic due to its unambiguous interpretation.
    \item The process carried out on spectrum samples strongly benefits from the processing gain provided by the fast Fourier transform (FFT) \cite{Lyons2004}. Although the transition between frequency and time domains can be performed quite efficiently, for simple energy detectors, it is not imperative to transform the signal into the frequency domain.
\end{itemize}
 With these considerations, we design a new separation technique, by reconstructing the  ROF-based solutions, for accurate extraction of parameters from intermittent PU transmission, while keeping the separation process as simple as possible to enable low-complexity detection. The main steps of the proposed algorithm are as follows:

\textbf{Initialization---energy vector}: The separation algorithm starts with the conversion of an $N$-sample signal time frame, i.e., $x=[x_{1},\dots x_{N}]$, into a vector of energy samples, $y_{n}=|x_{n}|^{2}$. Afterward, a moving average (MAV) of a small size $m_{\mathrm{init}} \ll N$ is used to  reduce noise variance initially. Because the recursive formulation of the $m$-sized MAV as  $\bar{y}_{n}=\bar{y}_{n-1}+\frac{1}{m}(x_{n}-x_{n-m})$ requires only one addition, one subtraction, and one division per sample, the formula is independent of the number of samples $N$, and the runtime complexity for each sample is constant, i.e., $\mathcal{O}(1)$. Thus, the complexity of the pre-processing preceding filtering is kept to a minimum.

\textbf{LOOP process---ROF}: In this step, the initially averaged energy vector $\bar{y}$ is iteratively filtered by consecutive erosion and dilation operations. A consistent increase in the size $k$ of  $movmin$ and $movmax$ filters allows finding the size $m_{\mathrm{sec}}$ for which the energy decrease in relation to the energy after previous filtration $e'_{k}=\frac{e_{k-1}-e_{k}}{e_{k-1}}$ remains the highest. The search for maximum value removes the requirement to set a threshold, as typical in earlier works. The resulting size of the filter $m_{\mathrm{sec}}$ is interpreted as the probable longest continuous signal duration in the analyzed frame, and is used in the second moving average,  $\bar{\bar{y}}$.

\textbf{Samples marking process---derivative evaluation}: The identification of signal and noise samples is based on the evaluation of the derivative of the double--averaged energy vector $y'_{n}=\bar{\bar{y}}_{n}-\bar{\bar{y}}_{n-1}$. The intervals in which the derivative has positive values with a width wider than the assumed threshold of minimal signal width $\lambda_{\mathrm{msw}}$ indicate signal samples. The threshold $\lambda_{\mathrm{msw}}$ can be interpreted as a resolution of the algorithm, i.e., the minimum detectable signal duration.
Due to the influence of the noise variance, the problem of maintaining the required interval continuity occurs. Positive intervals potentially indicating the signal, can be divided by single samples with non-positive values. Therefore, before assessing if the width of the positive interval meets the condition of $\lambda_{\mathrm{msw}}$, adjacent intervals separated by the single non-positive sample are combined. This simple step significantly improves the accuracy of the algorithm for weak signals. 
 \begin{algorithm}[H]
 \small
 \caption{Pseudo-code of the samples separation algorithm}
 \begin{algorithmic}[1]
 \renewcommand{\algorithmicrequire}{\textbf{Input:}}
 \renewcommand{\algorithmicensure}{\textbf{Output:}}
 \REQUIRE $x$, $m_{\textrm{init}}$, $\lambda_{\textrm{msw}}$
 \ENSURE  noise, signal
 \\ \textit{Initialisation} :
    \STATE{$y \!=\! (\textrm{abs}(x))^2 \!$, $\bar{y}\! =\! \textrm{mav}(y, m_{\textrm{init}})$, $e\! =\! \textrm{sum}(\bar{y})/\textrm{length}(\bar{y})$}
\\ \textit{Parallelized LOOP process:}
    \FOR{$k = 2 \textrm{~to length}(\bar{y})/2$}
		\STATE{$erosion = \textrm{movmin}(\bar{y},k)$}
		\STATE{$dilation = \textrm{movmax}(erosion,k)$}
	    \STATE{$e_{k} = \textrm{sum}(dilation)/\textrm{length}(dilation)$}
	    \STATE{$e'_{k} = (e_{k-1}-e_{k})/e_{k-1}$}
	\ENDFOR
	\STATE{$[value, index] = \textrm{max}(e'(2:end))$, $\bar{\bar{y}} = \textrm{mav}(\bar{y},index)$}
\\ \textit{Samples marking process:}
    \FOR {$j = 1 \textrm{~to length}(\bar{\bar{y}})-1$}
		\STATE{$y'_{j}$ = $\bar{\bar{y}}_{j}-\bar{\bar{y}}_{j+1}$}
	\ENDFOR
	\FOR{$l = 3 \textrm{~to length}(y')$}
		\IF{$y'_{l}>0$}
			\STATE{$y'_{l} = 1$}
			\IF {$y'_{l-2}>0$} 
				\STATE{$y'_{l-1} = 1$}
			\ENDIF
		\ELSE
			\STATE{$y'_{l}=0$}
		\ENDIF
	\ENDFOR
	\FOR {$i = 2 \textrm{~to length}(y')$}
		\IF {$y'_{i}$}
			\STATE{$y'_{i}=y'_{i-1}+1$}
		\ENDIF
	\ENDFOR
	\FOR{$n = 1 \textrm{\ to length}(y')$}
		\IF{$y'_{n}>\lambda_{\textrm{msw}}$}
			\STATE{$mark(n-\lambda_{\textrm{msw}}:n) = 1$}
		\ENDIF
	\ENDFOR
	\STATE{signal = $y \cdot mark$, $\textrm{noise} = y - \textrm{signal}$}
 \RETURN {noise, signal}
 \end{algorithmic} 
 \end{algorithm}

The above-presented solution results in a simple yet effective separation technique for the detection of packet-based PU signals. The pseudo-code of the proposed separation scheme is given in Algorithm 1 with notations: $m_{\textrm{init}}$-- initial size of the moving average, $\lambda_{\textrm{msw}}$-- minimal signal width, $x$-- received signal, $y$-- processed signal, $e$-- energy, $e'$-- energy decrease, and $y'$-- differential.

\section{Simulation Setup and Results}
\label{sec:simulation}

To assess the performance of the proposed algorithm, we have developed a simulation setup that supports random ON/OFF traffic behavior of PU. For background noise, the simulator uses radio-frequency (RF) noise I/Q traces sampled in a bandwidth of 5 MHz centered around 868 MHz, i.e., the first channel of the IEEE 802.15.4 standard. The RF noise traces are collected in an open university area using National Instrument USRP-2900. We recorded the average noise power of -110.7 dBm, which is normalized to 1 mW in the simulations.

\begin{figure}[!t]
	\centering
		\includegraphics[width=1\linewidth]{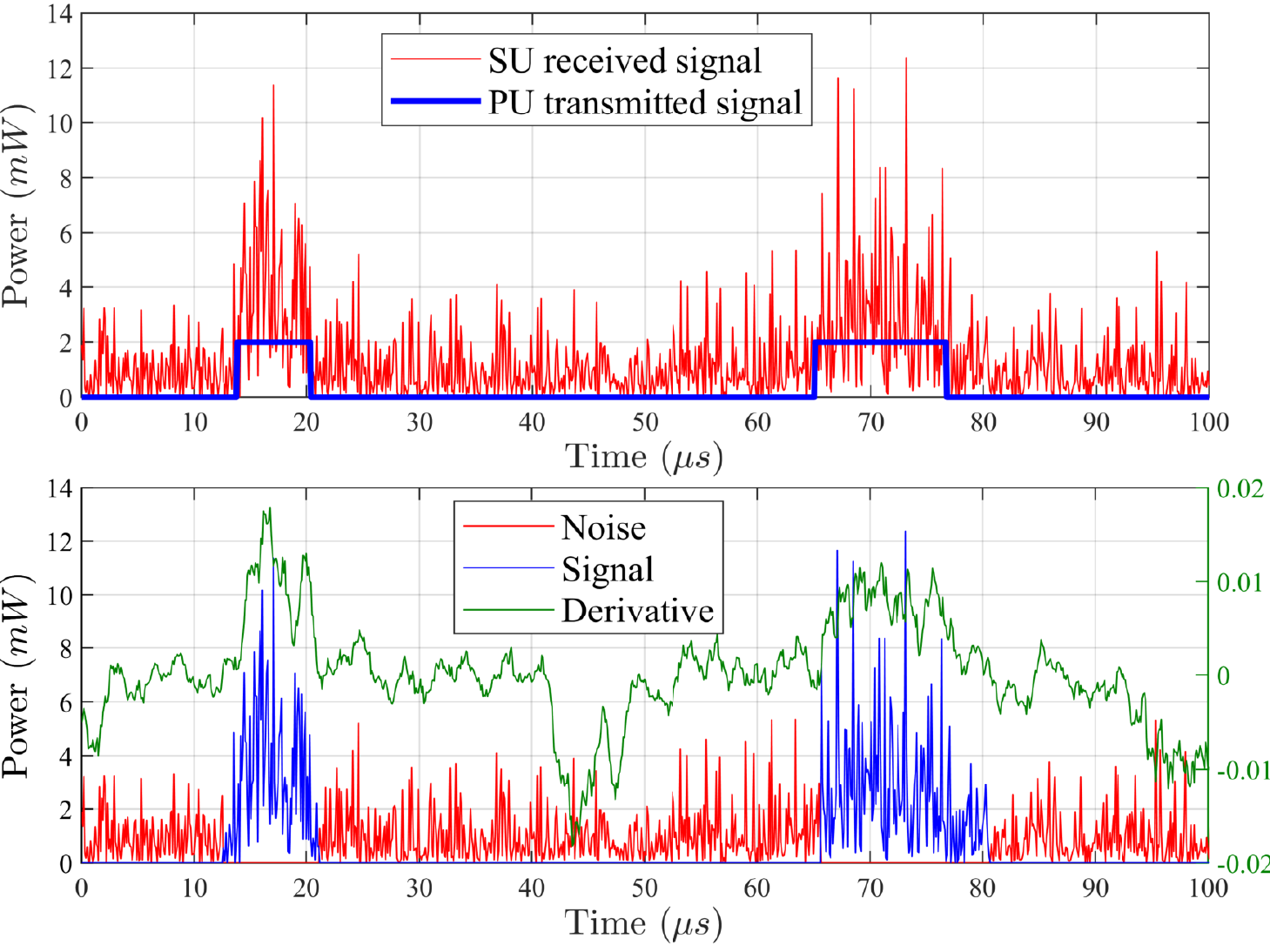}
	 \caption{An instance of analyzed time frame: (top)--the time frame with RF noise and PU pulse signals with randomly distributed durations. (bottom)--splitting of signal and noise samples based on the processed sign of the differential.}
	\label{fig:timeframe}
\end{figure}

In the simulations, the subject of analysis is non-overlapping time frames, where each frame contains 1024 noise samples to which a PU signal is added as a rectangular pulse as shown in Fig.~\ref{fig:timeframe}(top). Both the signal pulses and the following noise only duration are exponentially distributed with mean values being swept respectively from 10\% to 30\% and 90\% to 70\% of the time frame. 
The sample splitting is based on a positive or non-positive differential processed accordance to Algorithm~1, and depicted in Fig.~\ref{fig:timeframe}(bottom).

As the basic reference method, we study the estimation of primary user activity based on the idle/busy periods determined by using short spectrum sensing decisions \cite{Umebayashi2014, Iwata2019, Toma2020}. As the above method, however, requires noise floor information, which we obtain using the extended Forward Consecutive Mean Excision (FCME) algorithm with Welch FFT \cite{Umebayashi2014, Iwata2019}. For the simulation of the FCME-based algorithm, a 64-sample periodogram is adopted along with an energy detection window with a length of 5\% of the analyzed time frame and a false alarm probability of 1\%.

As the second reference method (operating in the time-domain), we adopt linear discriminant analysis (LDA). LDA is used in statistics and pattern recognition as a basic mathematical tool to separate two classes of objects, applied as Fisher discriminant function \cite{Chien2019}.

\begin{figure}[!t]
	\centering
		\includegraphics[width=0.90\linewidth]{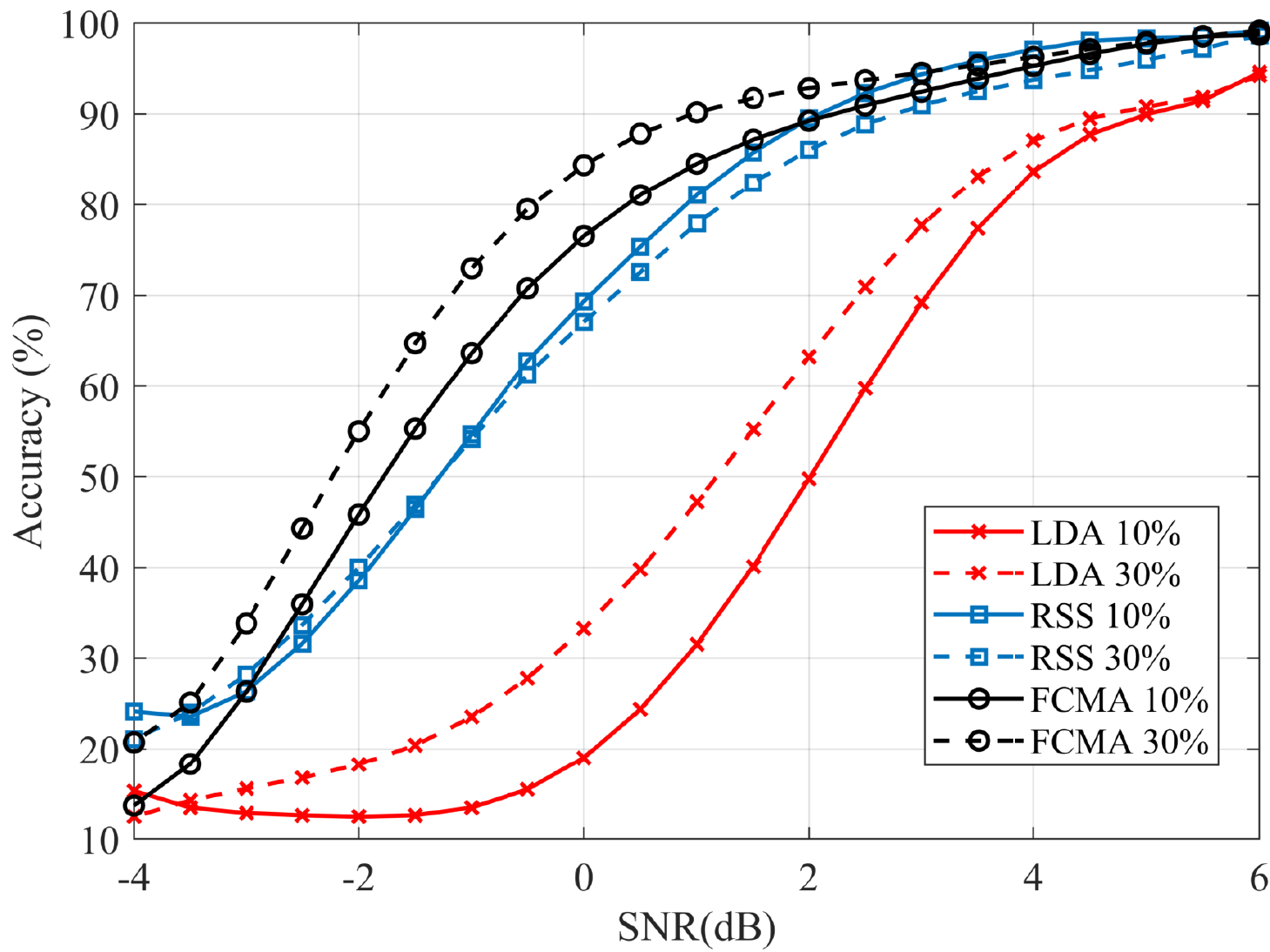}
	\caption{Signal samples marking efficiency as an average ratio of correctly recognized signal samples.}
	\label{fig:signalmark}
\end{figure}

The performance of the algorithms is measured by the average percentage of correctness in assigning samples to groups of signal-with-noise or noise-only samples, compared to a known pattern generated independently for each time frame. Fig.~\ref{fig:signalmark} shows the marking/separation accuracy for the signal group, indicating a decrease in the assignment accuracy with a decrease in SNR. 
The comparison indicates that the proposed ROF-based samples separation (RSS) shows a performance near to FCME-based solution and significantly higher than LDA (i.e., the reference time-domain method). Moreover, the accuracy of RSS remains almost independent of the mean occupancy of the time frame.

\begin{figure}[!t]
	\centering
		\includegraphics[width=0.90\linewidth]{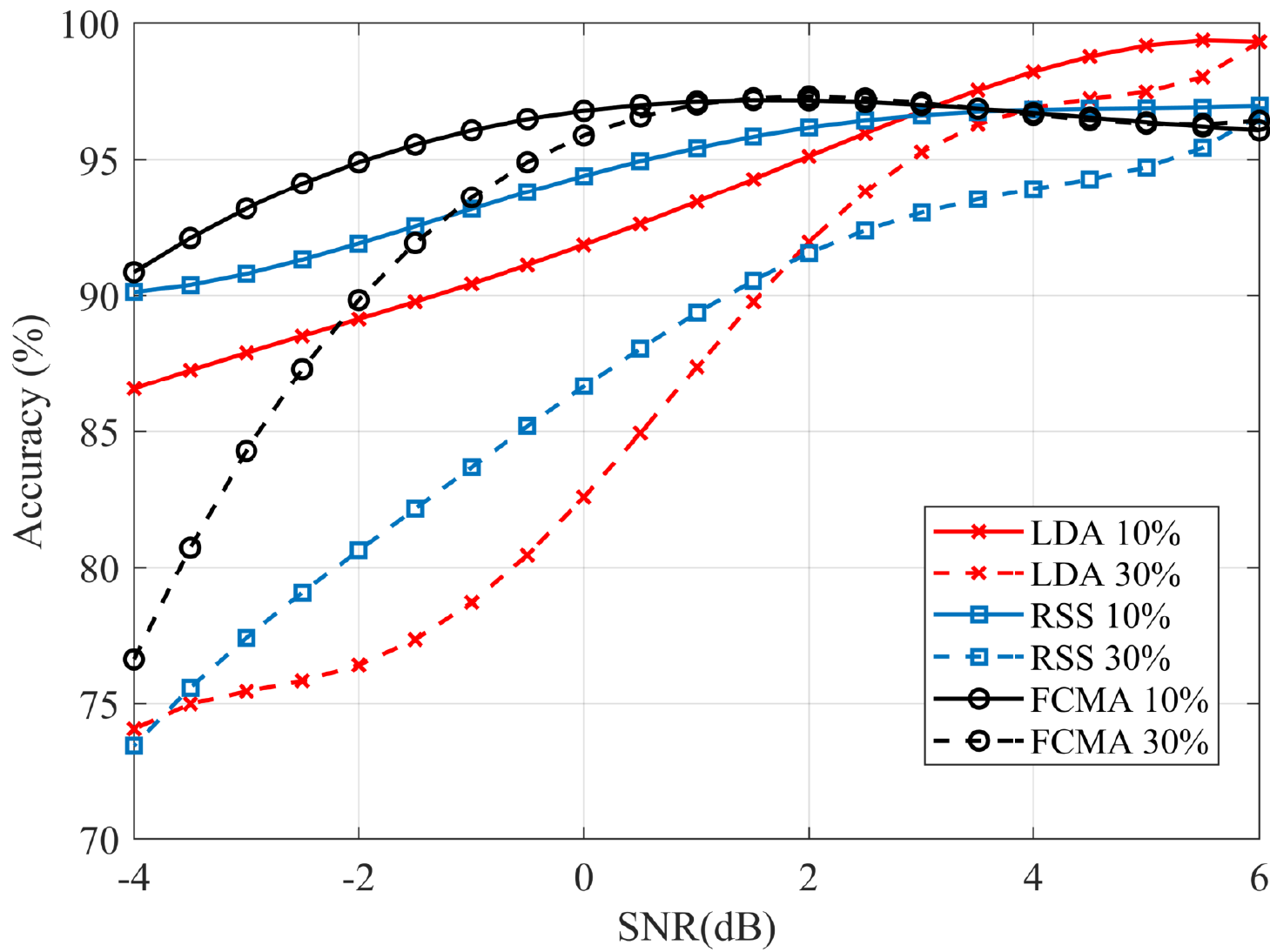}
	\caption{Total separation efficiency as an average ratio of correctly recognized signal and noise samples.}
	\label{fig:totalmark}
\end{figure}

The total marking accuracy, measured as an average ratio of correctly marked signal samples and noise samples, is shown in Fig.~\ref{fig:totalmark}. Note that the ability to correctly recognize the signal samples directly affects the overall separation efficiency. With a decrease in SNR, the efficiency of RSS decreases by the percentage of the signal presence in the frame, which is entirely recognized as noise in extreme cases.
Thus, the inaccuracy in RSS is mainly limited to error type false negative, i.e., less recognizable signal samples are assigned to the noise group, while the noise is rarely classified as a signal, leading to reduced false positive errors. At high SNR, the accuracy does not drop more than the assumed 5\% resolution threshold.

\begin{figure}[!t]
	\centering
		\includegraphics[width=0.90\linewidth]{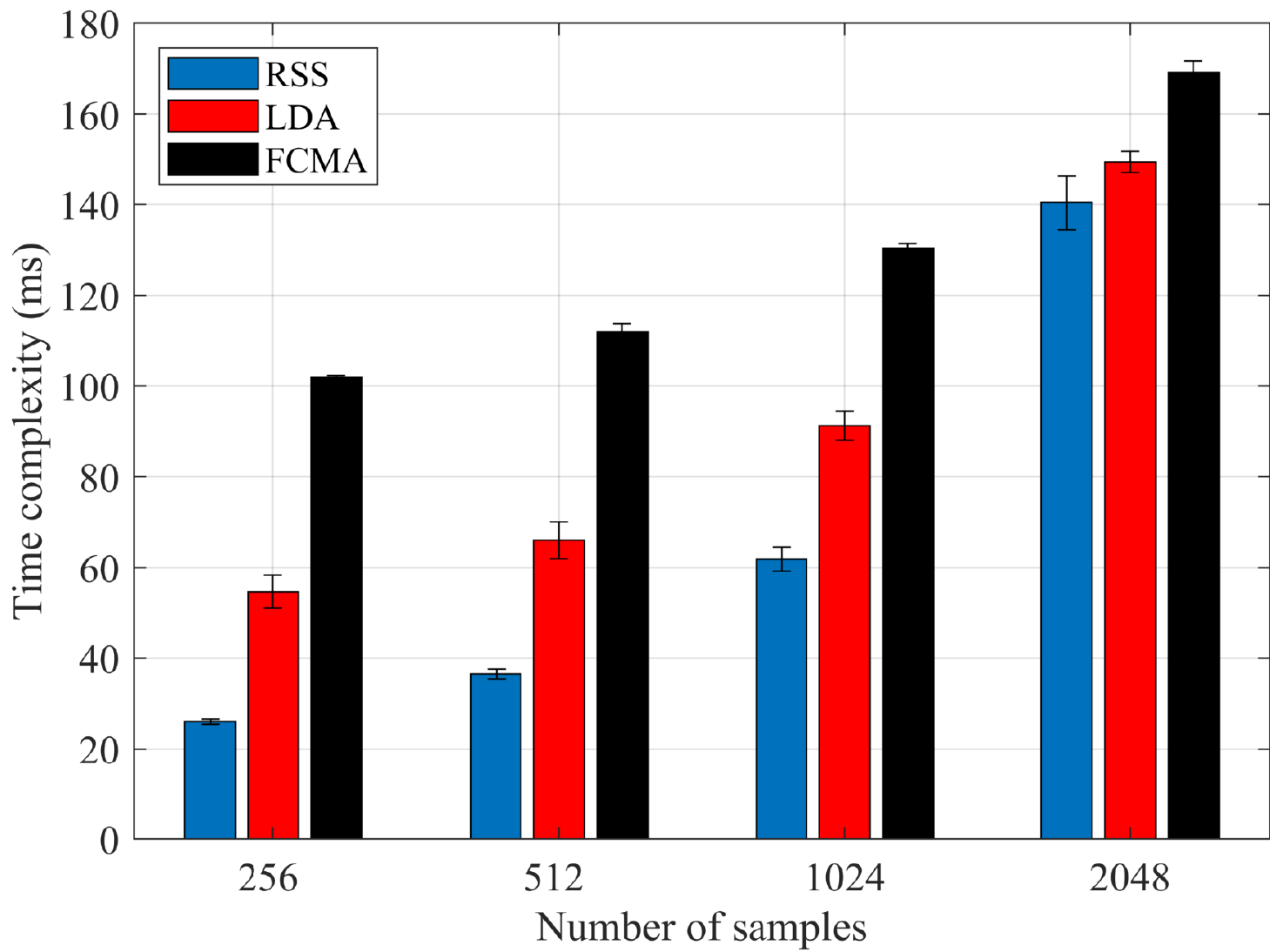}
	\caption{Time complexity as an average execution time of the separation process for a given number of samples.}
	\label{fig:complexity}
\end{figure}

Fig.~\ref{fig:complexity} compares the average execution time of the proposed RSS algorithm with the reference methods. The results were obtained by averaging the time over a thousand single-threaded function calls performed on a  quad-core 3.07  GHz  Intel  Xeon  W3550. Time analysis shows that the proposed RSS method exhibits significantly lower complexity, especially for a small number of samples, with respect to the reference solutions. These time differences are of significant importance as the sample separation remains only a supportive process for effective estimation of channel parameters and primary user detection.

\begin{figure}[!t]
	\centering
		\includegraphics[width=0.90\linewidth]{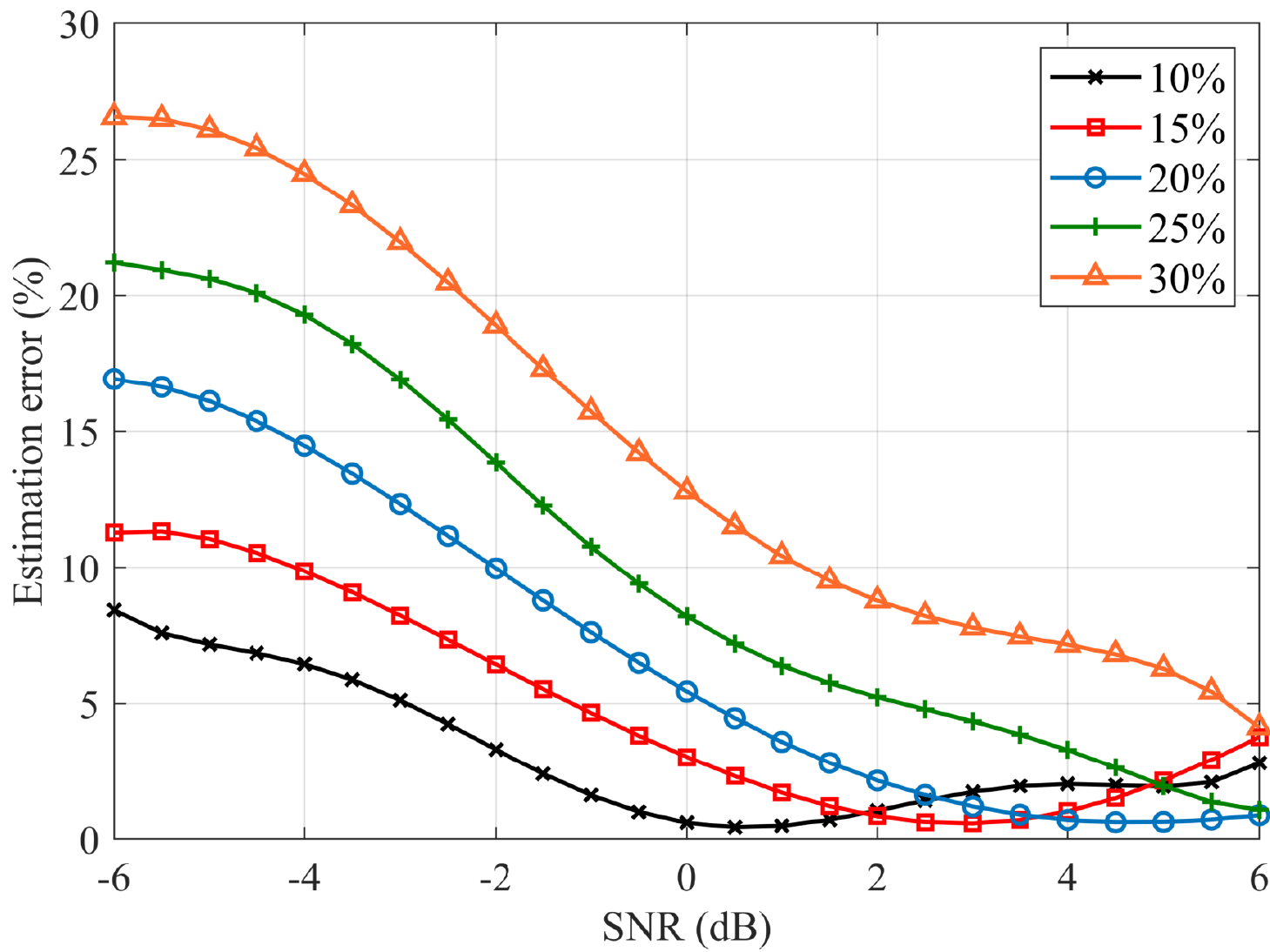}
	\caption{Accuracy of the mean channel occupancy estimation.}
	\label{fig:occupancy}
\end{figure}

The marking efficiency of signal-with-noise samples in the exemplary application can be directly used for estimating the average occupancy of the PU signal under test. As shown in Fig.~\ref{fig:occupancy}, it is  visible that for a low PU occupancy of 10--20\% with SNR above 0 dB, the proposed algorithm preserves high 1--2\% accuracy. In the case of increasing occupancy and weak signals, the algorithm noticeably loses efficiency. Therefore, for a secondary user subject to strong PU signals with a moderate activity factor, the proposed algorithm demonstrates to be a perfectly matched, easy to implement, and accurate separation technique.

\section{Conclusion}
\label{sec:conclusion}

Stimulated by the need for estimating several vital parameters to perform the energy detection of dynamic PU, we developed a new algorithm for accurate separation of signal and noise samples in the received signal time frame. The algorithm has its roots in rank order filtering-based spectral analysis that we reconstructed for low-complexity time-domain analysis of bursty signals. We evaluated the algorithm in terms of its accuracy to recognize signal samples, complete sample separation, time complexity and utility in channel occupancy estimation. The algorithm exhibits an accuracy of 87\% in separating samples even for weak signals with SNR close to 0 dB. For strong and narrow pulses, it provides up to 97\% of correct sample recognition and remains competitive for twice as complex solutions. The achieved accuracy, together with a simple design, makes the proposed solution a convenient basis for obtaining information required for effective energy detection.

\section*{Acknowledgment}
This work was supported by the Polish Ministry of Science and Higher Education within the status activity task 08/83/SBAD/4737 in 2019, and by the Swedish Knowledge Foundation under grant Research Profile NIIT.

\bibliographystyle{IEEEtran}
\bibliography{references}

\end{document}